\documentclass[a4paper,11pt]{article}
\pdfoutput=1 
\usepackage{pub} 
\usepackage[T1]{fontenc} 
\usepackage{subfigure}
\usepackage{amssymb}
\usepackage{abbreviations}
\usepackage{mathrsfs}
\usepackage{amsthm}
\usepackage{xspace}
\usepackage{lineno}
\usepackage{multirow}
\usepackage{appendix}
\usepackage{float}

\title{\boldmath Direct optimisation of the discovery significance when
training neural networks to search for new physics in particle
colliders}

 \author{Adam~Elwood}
 \author{and Dirk~Kr\"ucker}
 \affiliation{Deutsches Elektronen-Synchrotron,\\Notkestr.\ 85, 22607 Hamburg, Germany}
\emailAdd{adam.elwood@desy.de}
\emailAdd{dirk.kruecker@desy.de}

\abstract{We introduce two new loss functions designed
to directly optimise the statistical significance of the expected
number of signal events when training neural networks to classify
events as signal or background in the scenario of a search for new
physics at a particle collider. The loss functions are designed to
directly maximise commonly used estimates of the statistical
significance, $s/\sqrt{s+b}$, and the Asimov estimate, $Z_A$. We
consider their use in a toy SUSY search with 30~fb$^{-1}$ of 14~TeV data
collected at the LHC. In the case that the search for the SUSY model is dominated by
systematic uncertainties, it is found that the loss function based on $Z_A$ can
outperform the binary cross entropy in defining an optimal search
region.}

\begin{document} 
\maketitle
\flushbottom
%\linenumbers

\section{Introduction} \label{sec:intro}

Data analysis in High Energy Physics (HEP) is a genuine multivariate
classification
problem. In particle colliders, the results of collisions that are
recorded by detectors, such as ATLAS and CMS at the LHC, are 
reconstructed to determine the subatomic particles produced in each
collision. In searches for new physics, a
multivariate analysis is carried out with these quantities to distinguish between the
Standard Model (SM) background processes and potential signal from physical
processes that are not predicted within the SM. 

Traditional searches for new physics, such as 
supersymmetry~\cite{Ramond:1971gb,Golfand:1971iw,Neveu:1971rx,Volkov:1972jx,Wess:1973kz,Wess:1974tw,Fayet:1974pd,Nilles:1983ge}, approach
this by designing discriminative {\it high level} variables, based on
the result of the event reconstruction. These variables usually
exploit differences in energy scale or topology between the signal and
background processes. The design of these variables requires
significant prior physics knowledge. 
Additionally, 
this approach may not take into account all of the
subtle differences between signal and background events.

There have been significant advances in the power and viability of
using neural networks for such discrimination problems in recent
years, mainly driven by the accessibility of large datasets and
advances in computing power. 
Deep neural networks have the promise to
approach the background and signal classification problem from
{\it low level} reconstructed quantities. They are then able to find
sophisticated ways to discriminate between signal and background,
finding details above and beyond those contained in the high level variables
\cite{Baldi:2014kfa}.

When a new physics search is framed as a signal and background
classification problem, the search is actually designed to maximise
the statistical significance of the signal sample over the background
sample, rather than obtaining the best classification accuracy or
receiver operating characteristic (ROC). This translates to a maximisation of the
correct classification of signal events, while minimising the
incorrect classification of background events. In this case, the
classification of signal events as background events is tolerable,
providing the overall purity of the signal classification is maintained. Many of the approaches for optimisation of
classification problems with neural networks, such as the minimisation
of the cross entropy, 
give the same weight to the correct classification of signal events as to the correct classification of background events. 
In this paper, we consider an alternative
optimisation approach that directly maximises the statistical
significance of the selected signal.
This gives more priority to the purity of the signal classification
than the purity of the background classification.

\section{An example search for a third-generation supersymmetric quark partner}\label{sec:stop}

As a physics example we consider a toy search for supersymmetric top
quarks at the LHC. The search is designed for the case of direct
top squark pair production with subsequent decay of each squark
into a top quark and the lightest supersymmetric particle (LSP). 
We assume that the top squark and the LSP are the only SUSY particles
that have low enough
masses to be accessible at the LHC. In a real search, limits are
typically set as a function of the top squark and the neutralino
masses.  In the context of SUSY searches
models are often categorised in two different ways. In the cases that the top
squark is much heavier than the LSP, the SM decay products are
produced with significant energy, making them easy to observe in the
detector. These models are known as {\it uncompressed}. In the case
that the difference in mass between the top squark and LSP is small,
the SM decay products are produced close to rest, making the signature
much more difficult to distinguish from the background. These models
are known as {\it compressed}.

For our toy study, we assume two sets of mass parameters that
are expected to be on the border of discovery with about 30\fbinv of
14\TeV data collected at the LHC. We consider
an uncompressed
point with the top squark mass fixed at 900\GeV and
the neutralino mass at 100\GeV. We additionally consider a compressed mass point
with a top squark mass of 600\GeV and a
neutralino mass of 400\GeV,
where the mass splitting is at the order of the top mass. These two
points present different challenges, the uncompressed point has a
low cross section, meaning few signal events are expected to be
produced. The compressed point has a higher cross section
but is harder to
distinguish from the background.
As background, we consider only the
dominant process of top-antitop (\ttbar) production. This background is several
orders of magnitude higher than the signal, and the distributions of
signal and background variables are quite similar, due to their
comparable kinematics. 

A leading-order simulation is sufficient for our purpose and we only
take into account next-to-leading order results for the total cross
sections of stop signal~\cite{Borschensky:2014cia} and \ttbar
background~\cite{Campbell:2010ff,Nadolsky:2008zw}. 
Cross sections of 17.6~fb for the uncompressed model, 228~fb for the
compressed model and 844~fb for the \ttbar background are used.     
{\sc Pythia8}~\cite{Sjostrand:2014zea,Sjostrand:2006za} is used for the
event simulation and {\sc Delphes3}~\cite{deFavereau:2013fsa} to
model the detector response using the Delphes model of the CMS
detector. Jets are clustered with anti-$k_T$~\cite{Cacciari:2008gp} with a cone
parameter of $R=0.4$ and we use {\it lepton} as a generic term for electrons
and muons. For simplicity, we do not consider tau leptons since their
experimental reconstruction is more complex.

In order to reduce the training times, while retaining most of the
events of interest, signal and background events
are pre-selected.  We require at least one lepton with transverse
momentum $\pt > 30\GeV$ within a pseudorapidity of $|\eta| < 2.4$.
Each event must contain at least four jets with $\pt > 40\GeV$, where
the highest-\pt (leading) jet is required to have
$\pt>80\GeV$, and the sub-leading jet $\pt>60\GeV$. At least one of
the jets must be tagged as originating from a bottom quark.  The
missing energy perpendicular to the beam direction (the negative vector sum of the momenta of all reconstructed particles), \ETslash, is
required to be above 200\GeV, and the scalar sum over the transverse
momenta of all preselected jets, \HT, to be above 300\GeV.  After this
preselection, the remaining background is still several orders of
magnitude higher than the investigated signal.  This setup is similar
to the one used in~\cite{Sahin:2016qgg} where more details can be
found.  
The selected sample, which is used for training and tuning,
consists of $1.4\times 10^6$ events with a sample composition of $50\%$
signal and $50\%$ background events. An independent sample of
$6\times 10^5$ events is used for
testing and employed only in the evaluation step.

We consider both high level and low level variables when training the
networks. The
low level variables consist of basic properties ($E, \pt, \phi$ and
$\eta$) of the reconstructed physics objects, i.e.\ of the three
leading jets and the leading lepton. 
In addition, the multiplicities
of jets  ($n_{\mathrm{jet}}$) and b-quark jets
($n_{\botq-\mathrm{jet}}$) are considered. As high level quantities we
consider \ETslash as well as \HT.  In our SUSY
models, we expect \ETslash from the LSP, which is expected to
be neutral and weakly interacting and will therefore not be detected.
As SUSY particles are heavy, we also expect a large amount of energy
in the detector leading to large \HT. 

We also consider more sophisticated high level variables that are
commonly used in SUSY searches. The
transverse mass, defined as $\MT = \sqrt{ 2\,
p_{\mathrm{T},\lep}\,  \ETslash\,  (1 - \cos \Dp (\lep,\ETslash))}$,
where $\Dp (\lep,\ETslash)$ is the azimuthal angle between the lepton
and the \ETslash vector, can be used to suppress the background from W
boson production, as \MT of leptonic W decay events does not exceed
the W mass. An important background comes from \ttbar events in which both
top decays produce leptons and one of the leptons is not properly
reconstructed. In this case the lost lepton mimics large missing energy
from the LSP. The \MTtW variable~\cite{MT2W} is constructed exploiting
the knowledge of the \ttbar-decay kinematics to separate such events.
Since top squark production is a high-mass process with large missing
energy, it results in a higher value of \MTtW than the background.
A summary of all low-level
and high-level variables used as input features to the neural networks
detailed in this paper is given in Table~\ref{tab:varsets}.
\begin{table}[!h] \caption{\label{tab:varsets} Summary of all
  low level and high level variables used in this analysis as
  described in the text. } \vspace{2em} \centering
  \begin{tabular}{cc} \hline low level&high level \\ \hline
    $\vec{p}_{\lep}$            &  $\MT$    \\ 
    $\vec{p}_{jet(1,2,3)}$ &  $\MTtW$ \\
     $n_{jet}$                                 &$\ETslash$ \\ 
     $n_{\botq\,jet}$                       & $\HT$\\ 

\hline \end{tabular} \end{table}

\section{Performance measures of machine learning techniques applied to particle
physics searches} \label{sec:perform}

In the case of supervised learning, the quality of a binary classifier
is typically described by some kind of measure that quantifies how
well the machine learning algorithm separates the two distinct
classes. On a sample of test data, where we know the true class
labels, there are 2x2 categories formed by the true and the estimated
labels. The matrix of entries in these categories is known as
confusion matrix and the relative amount of test data in these
categories can be used to quantify the performance of a machine
learning algorithm. Typical performance measures are the {\sl accuracy}, the
percentage of true positive and true negative labels, or the AUC, the
area under the ROC curve~\cite{Bradley:1997}. They can be used as
evaluation metrics during the training of a neural network, while the
training itself uses some kind of differentiable loss function to
allow for backpropagation. The cross entropy is the typical loss
function chosen for binary classification (see for example~\cite{murphy12}). 

In~\cite{Sahin:2016qgg}, it had been argued that optimising for accuracy or AUC
is not appropriate for a HEP search. The true positive with respect to
the false positive labels are more relevant in this case. 
The number of false negative labels, i.e. the contamination of the
background classification, is less important.
In a physics search, background and
signal often overlap in a large part of the available phase space and
the performance of a classifier here can be suboptimal. The
problem that must be solved is not to label background and signal
correctly, but to find an area in phase space where the signal
dominates significantly over the background. Then positive labels
should only be given to this subset of signal events. Sidestepping from the logic of
supervised learning and binary classification, we use the neural
network to define a (not necessarily simply connected) subspace in our
feature space where the signal events dominates in a statistically well
defined way. This defines a single bin where we count
signal and background events.

Given these considerations, training the
neural network with cross entropy as a loss function will not
necessarily result in
an optimal search region. The loss function must be modified in an
appropriate way.
Starting with the concept that the trained classifier will cut out an
area in phase space where a certain amount of signal and background
events are observed, we aim for the best region such that the
discovery significance for Poisson distributed events becomes maximal.
We want to define an optimal search before we observe the data and
maximize the expected discovery significance. The exact numerical
calculation of the statistical significance may
become computationally costly but a well performing estimate for the
discovery significance, known as the \textsl{Asimov} estimate, has been
given in  \cite{CCGV2011}.  For the case of Poisson distributed
background ($b$) and signal ($s$) events with background uncertainty
$\sigma_b$, the approximated median discovery significance becomes:
\begin{equation} \label{eq:asimov}
Z_{A}=\left[2\left((s+b)\ln\left[\frac{(s+b)(b+\bsvar)}{b^2+(s+b)\bsvar}\right]-\frac{b^2}{\bsvar}\ln\left[1+\frac{\bsvar s}{b(b+\bsvar)}\right]\right)\right]^{1/2}.
\end{equation}
In the case where the background is known exactly ($\sigma_{b}=0$)
this simplifies to:
\begin{equation} 
 \label{eq:asimov2}
Z_{A}(\sigma_{b}=0)=\sqrt{ 2\left((s+b)\ln (1+s/b) - s \right)},
\end{equation}
and for the case where $s,\bsvar\ll b$ we are left with
$s/\sqrt{b+\bsvar}$. The systematic uncertainty $\bsvar$ is
assumed to be proportional to $b$ and given as a percentage on the
true background events in the following sections. 

In addition, we use in the following $s/\sqrt{s+b}$, which can be understood as the exclusion significance
in the large sample limit.  

Unlike accuracy or cross entropy, these expressions depend on the
absolute number of events. It will therefore be necessary to consider the
proper event weights to describe the physical event counts.  Where we
give errors on the significance, we use error propagation to the
significance approximations and assume Poisson errors for the number
of selected events before applying event weights.

\section{Loss functions to directly optimise neural networks for discovery significance} \label{sec:NN}

The new loss functions designed to directly select an area of phase
space that is more appropriate to background and signal discrimination
in a physics analysis are described in this section. We consider two
commonly used approximations of statistical significance,
$s/\sqrt{s+b}$ and the Asimov estimate, outlined in
Sec.~\ref{sec:perform}. 
The two estimates
are defined within the context of a training
batch, where $s$ corresponds to the number of correctly classified
signal events and $b$ corresponds to the number of background events
classified as signal. 
To ensure differentiability of the loss
function,
these classifications cannot be 
discrete. The output layer
of the network is therefore chosen to be a single neuron with a
sigmoid activation function. 
When used as a performance metric,  
the prediction of a signal event is
defined by cases in which the output neuron has values above $0.5$, with values below
classified as background events. 
When used as a loss function differentiability must be ensured so,
the values of $s$ and $b$ are defined as follows:
\begin{equation}
  s=W_s\sum_{i}^{N_{batch}} y^{pred}_i\times y^{true}_i ,
\end{equation}
\begin{equation}
  b=W_b\sum_{i}^{N_{batch}} y^{pred}_i\times (1-y^{true}_i) ,
\end{equation}
where $N_{batch}$ is the number of training events in each batch,
$y^{pred}_i$ is the value of the final sigmoid output for event $i$ in
the batch and $y^{true}_i$ is its true value, 1 for signal and 0 for
background. $W_s$ and $W_b$ are the physical weights of the signal and
background samples that scale the total number of signal and background
events in each batch up to the number expected to be observed given a
certain luminosity, $L$, their cross sections, $\sigma_{signal}$ and
$\sigma_{bkgd}$, and the efficiency of acceptance of events in the
preselection, $\epsilon$. They are defined as:
\begin{equation}
  W_s=L\sigma_{signal}\epsilon/N^{true}_{signal},
\end{equation}
\begin{equation}
  W_b=L\sigma_{bkgd}\epsilon/N^{true}_{bkgd},
\end{equation}
where $N^{true}_{signal}$ and $N^{true}_{bkgd}$ are the number of true signal
and background events in the batch respectively (i.e.
$N^{true}_{signal}+N^{true}_{bkgd}=N_{batch}$). The above definitions ensure
that $s$ and $b$ have continuous values as the weights of the network
update and the value of $y^{pred}$ changes. It can be seen that in the
limit of an absolute certainty in the value of $y^{pred}$, either 1 or
0, $s$ and $b$ converge to a discrete definition.

Given these definitions of $s$ and $b$, two loss functions are defined
as the inverse of the two approximations of significance
that are considered. For the $s/\sqrt{s+b}$ approximation the loss
function is
defined as:
\begin{equation}
  \ell_{s/\sqrt{s+b}} = (s+b)/s^2,
\end{equation}
in which the approximation is inverted, to turn the
maximisation of the significance into a minimisation problem, and
squared, to 
reduce the number of expensive computations required. The loss
function to maximise the Asimov estimate is defined as
\begin{equation}
  \ell_{Asimov} = 1/Z_A,
\end{equation}
where $Z_A$ is defined in~\ref{eq:asimov}.

\begin{figure}[t]
\centering
\subfigure[]{
  \includegraphics[width=0.49\textwidth]{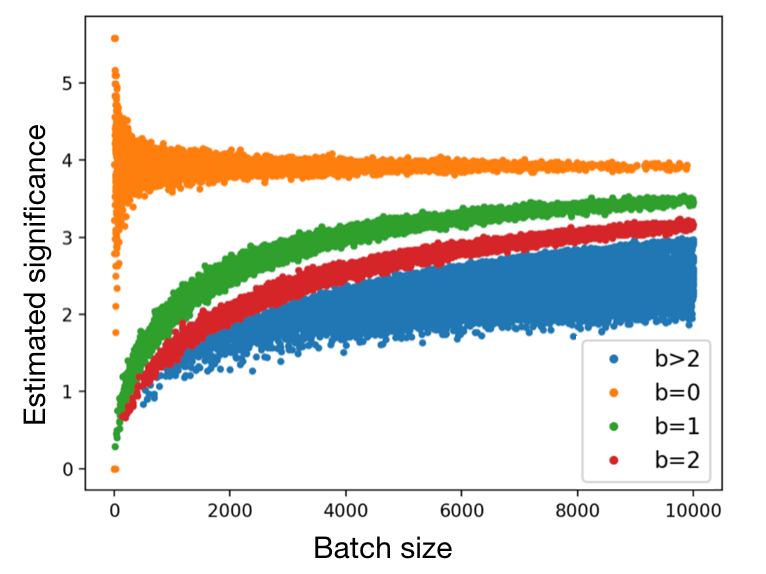}
}~
\subfigure[]{
  \includegraphics[width=0.51\textwidth]{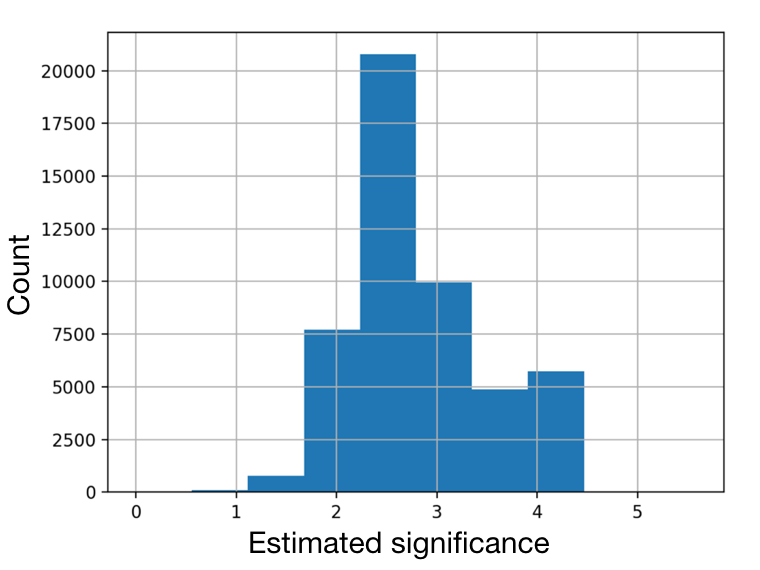}
}
  \caption{The $s/\sqrt{s+b}$ approximation of significance for a
  variety of batch sizes. Each batch is randomly sampled from the full
  set of events, with 10 points plotted for each batch size
  considered. A histogram of the significances from all of these
  attempted batch sizes is plotted on the right (b). 
  }
\label{fig:batchSize}
\end{figure}

The degree to which $s$ and $b$ are approximated well in each batch
depends strongly on the batch size. If the batch is too small and the
network is well trained to reject background, it is quite feasible
that for a particular batch, $b=0$.
To avoid strong statistical fluctuations of the estimated
significance, a sufficient number of signal and background events is necessary.
This problem can be seen in Fig.~\ref{fig:batchSize}, in
which an estimate of significance ($s/\sqrt{s+b}$) is plotted for a series of
different batch sizes. For batch sizes in the range from 2 to 10000,
each batch is randomly sampled 10 times from the full dataset of
signal and background events. For each of these instances a point is
plotted on the plot on the left and
added to the histogram on the right. The significance here is plotted
after a network has been trained for 20 epochs, optimising with
$\ell_{s/\sqrt{s+b}}$. Calculating the significance with the full
dataset gives a value of $2.5\,\sigma$, which qualitatively agrees with
the distribution
in the histogram. However, in the cases that a low number of
background events are classified as signal, the
significance can fluctuate to a higher value, giving an erroneous
estimate. To reduce the chance of this happening, a large batch size
is desirable. It was found empirically that in this case a batch size of
4096 worked well. The option of dynamically varying
the batch size was also considered, starting off with a small batch
and increasing it with subsequent epochs to improve the accuracy of
the significance prediction. It was found that this did not
produce a dramatic improvement above directly choosing an adequately
large batch size for the full training of the network. However, it may still be
a useful technique for getting a good result when hyperparameter
optimisation of the batch size is not feasible.

When optimising with $\ell_{Asimov}$, it was found that it would take many
iterations of weight updates before the loss started minimising. In
many cases the loss would not reduce below its initial value for a
large number of epochs. This is due to the gradient of the Asimov
estimate being small in the cases that $s$ is small with respect to
$\sigma_b$.
This can be resolved
by carrying out a pretraining with $\ell_{s/\sqrt{s+b}}$, which has a
higher gradient than $\ell_{Asimov}$ when $s/b$ is low. We find that a
pretraining of $\sim5$ epochs is typically enough to increase $s/b$ to
a reasonable level,
$\ell_{Asimov}$ can then be used to finish off the optimisation of the
network.

\section{Results} \label{sec:results}

To demonstrate the performance of the loss functions described in
Sec.~\ref{sec:NN}, they are used to optimise a variety of simple neural
networks that aim to discriminate signal from background in the toy
SUSY analysis described in Sec.~\ref{sec:stop}. All of the high and
low level variables summarised in
Tab.~\ref{tab:varsets} are scaled so as to have a mean of 0 and a
standard deviation of 1 and subsequently used as input features to the neural
networks.

Dense networks with three distinct topologies are trained, one with a
single hidden layer of 23 neurons, one with three hidden layers of 46 
neurons and one with five hidden layers of 23 neurons. They all
give the same qualitative results and have a similar performance
on our simple toy problem.
The results shown in this section are evaluated from the network with
a single hidden layer, as it had the least sensitivity to overtraining
and adequately demonstrates the performance. 

\begin{figure}[t]
\centering
\subfigure[Loss evolution]{
  \includegraphics[width=0.5\textwidth]{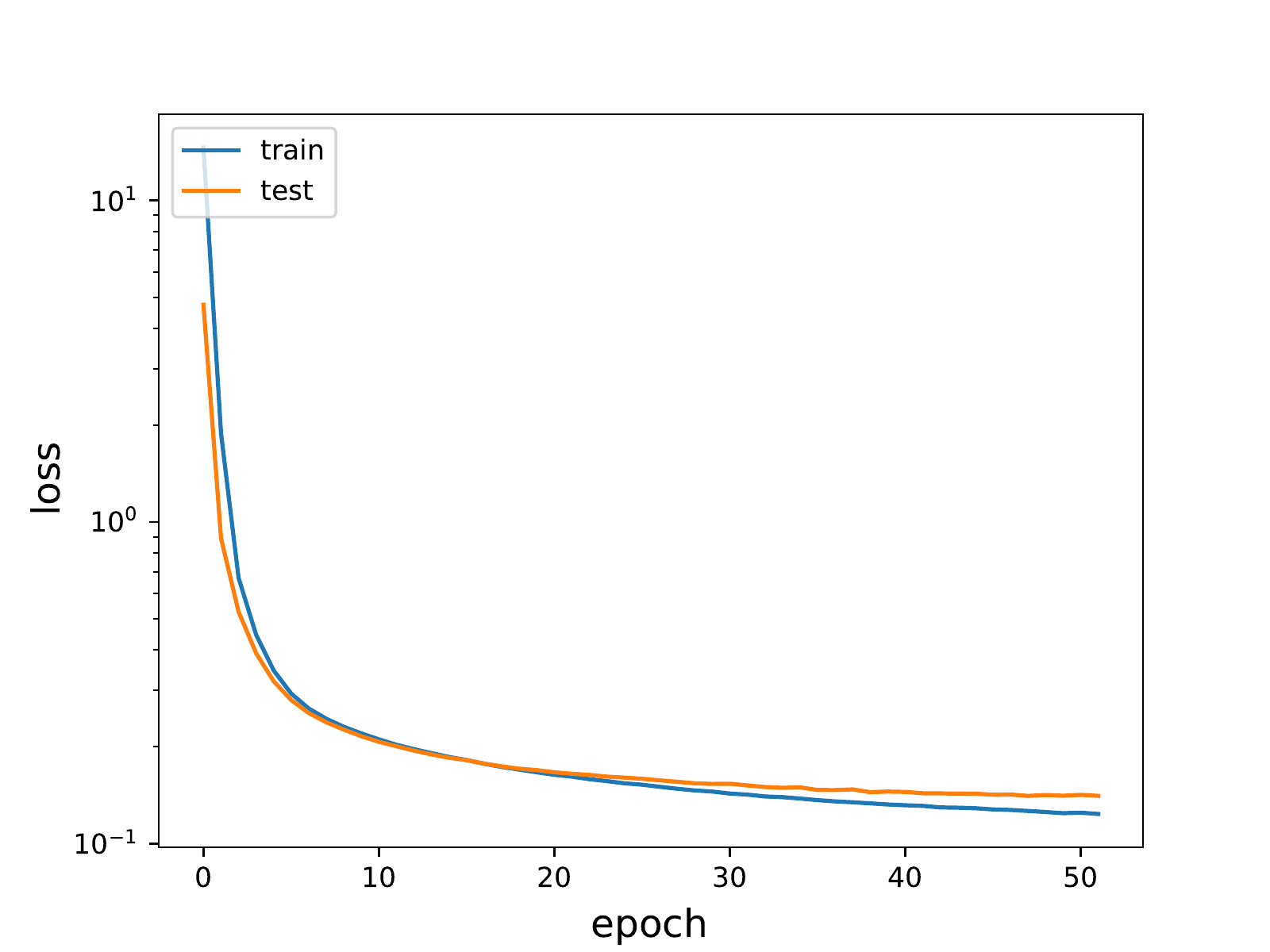}
}~
\subfigure[Accuracy evolution]{
\includegraphics[width=0.5\textwidth]{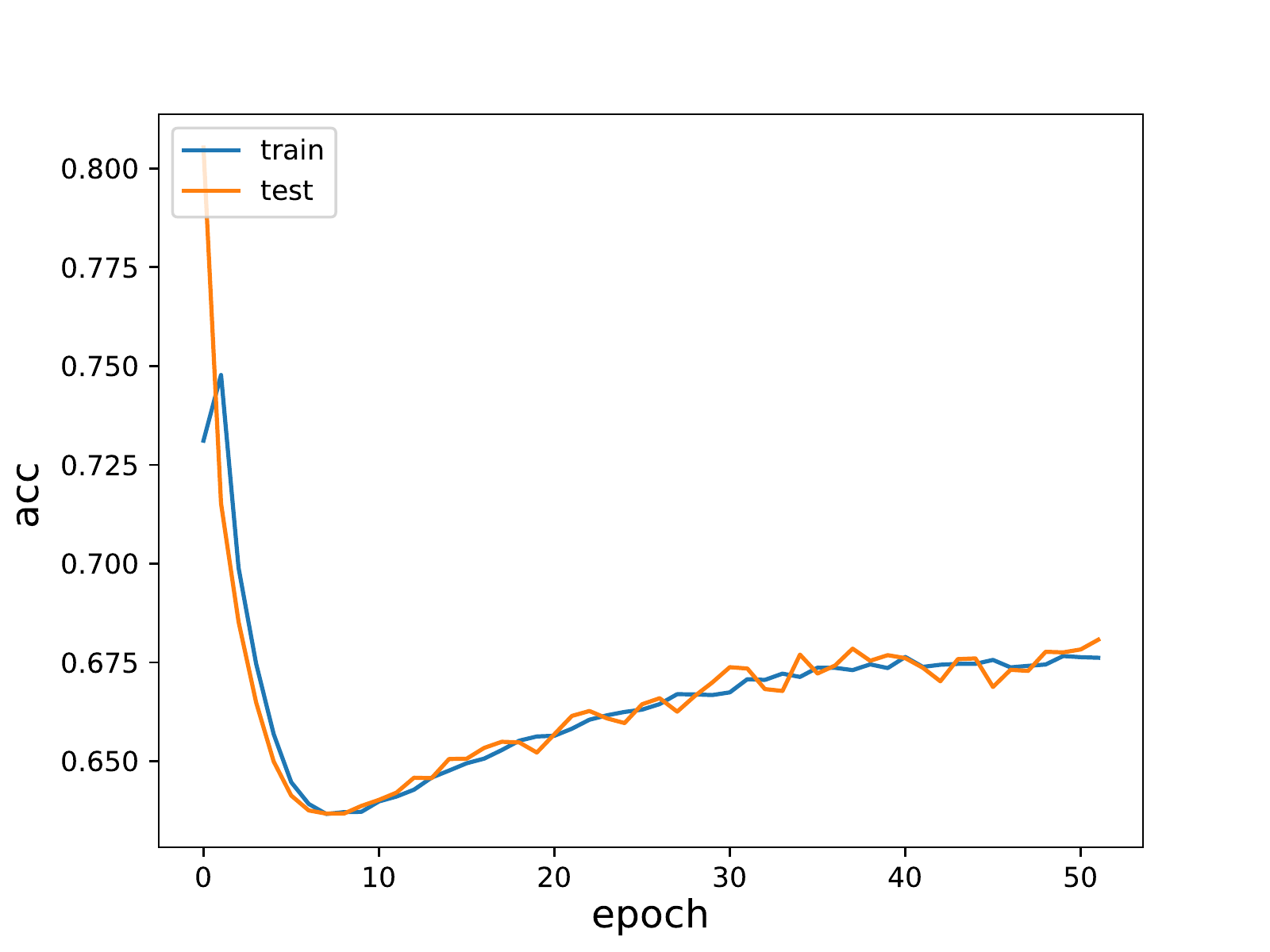}
}
  \caption{The evolution of the loss function (left) and accuracy of
  the model (right) for a single hidden layer neural network optimised
  with $\ell_{s/\sqrt{s+b}}$.}
  \label{fig:lossaccevolution}
\end{figure}

\begin{figure}[t]
\centering
  \subfigure[Binary cross entropy]{
    \includegraphics[width=0.5\textwidth]{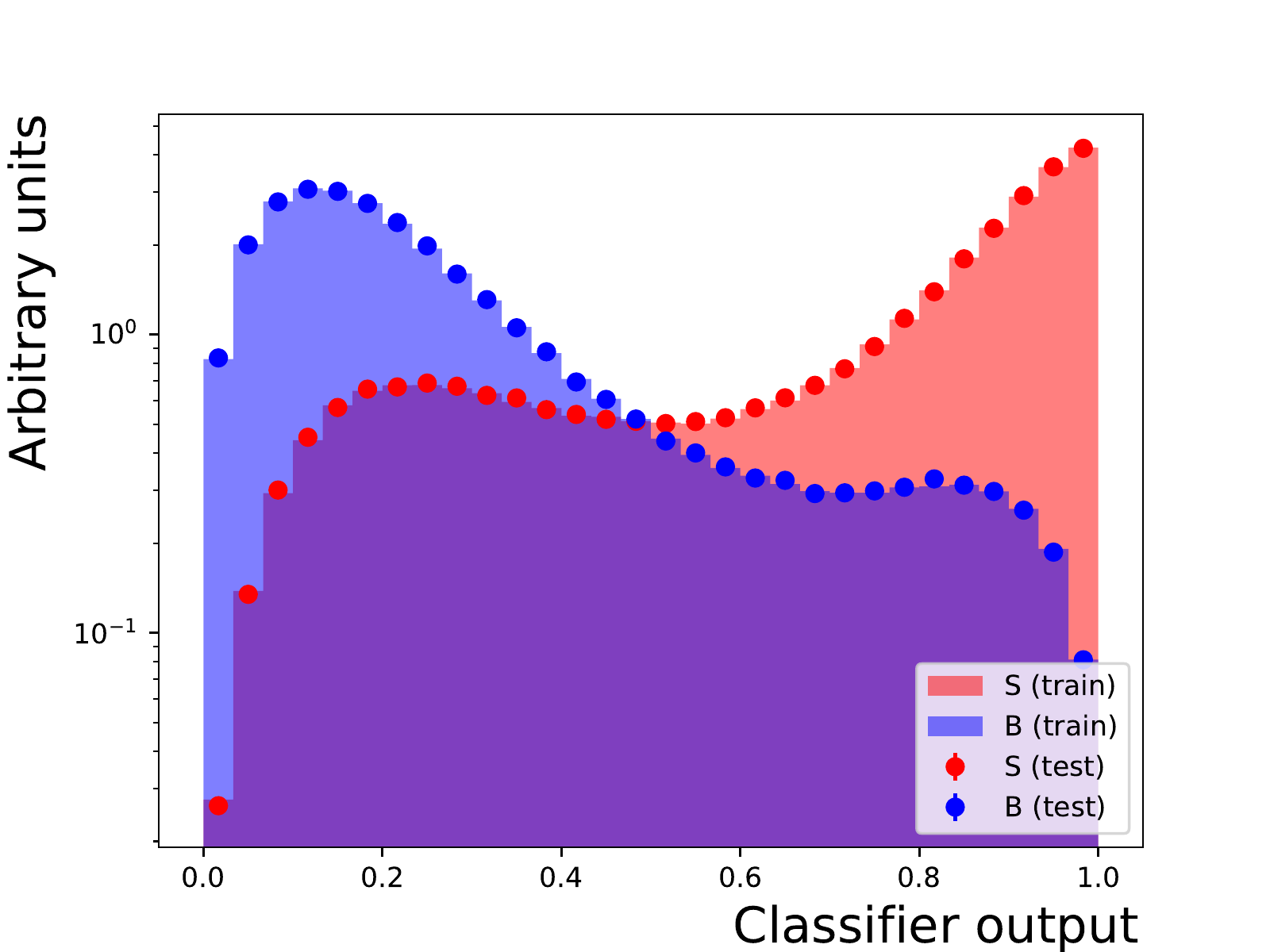}~
  }~
  \subfigure[$\ell_{s/\sqrt{s+b}}$]{
\includegraphics[width=0.5\textwidth]{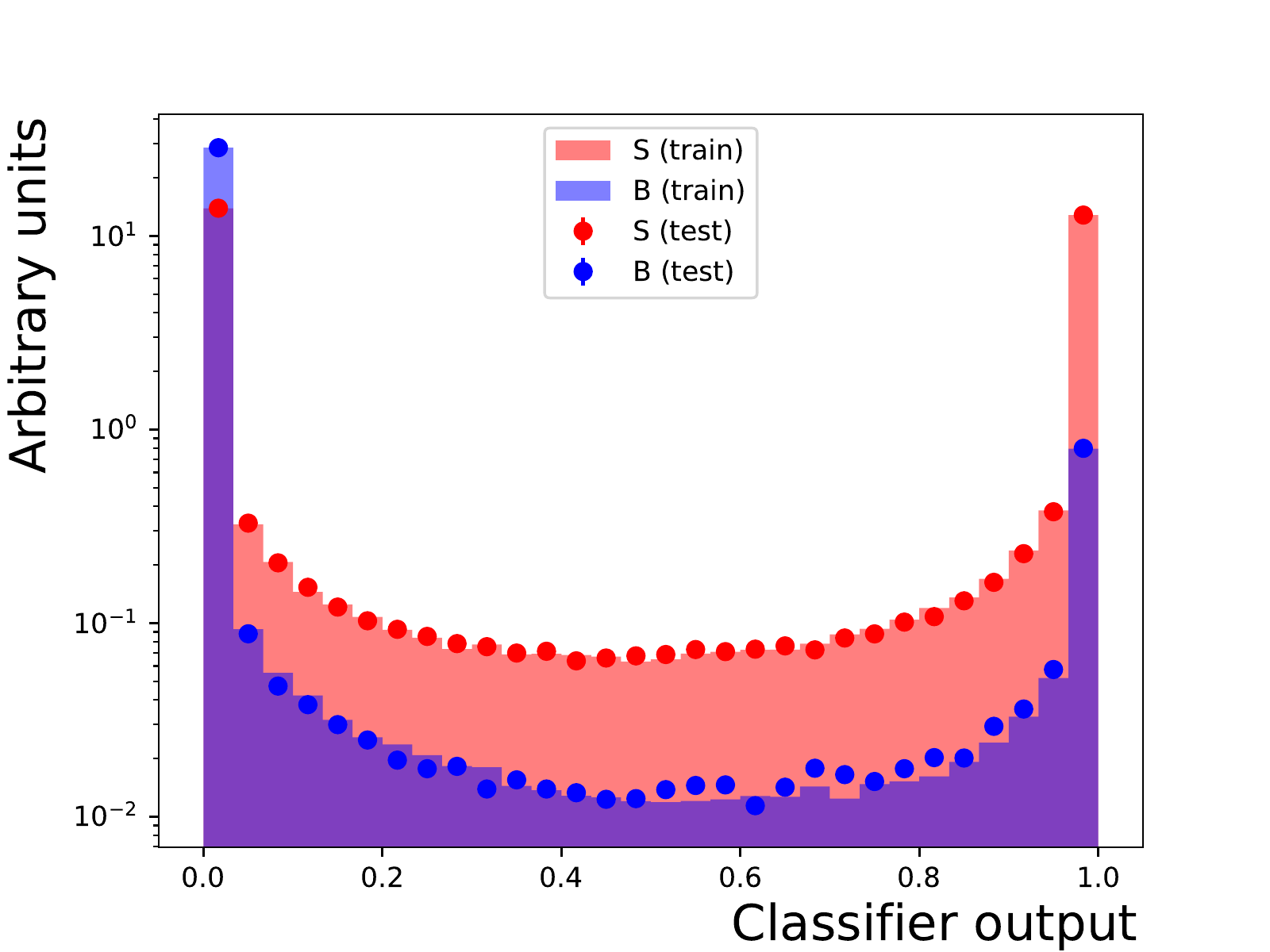}
  }\\
  \subfigure[$\ell_{Asimov}$]{
\includegraphics[width=0.5\textwidth]{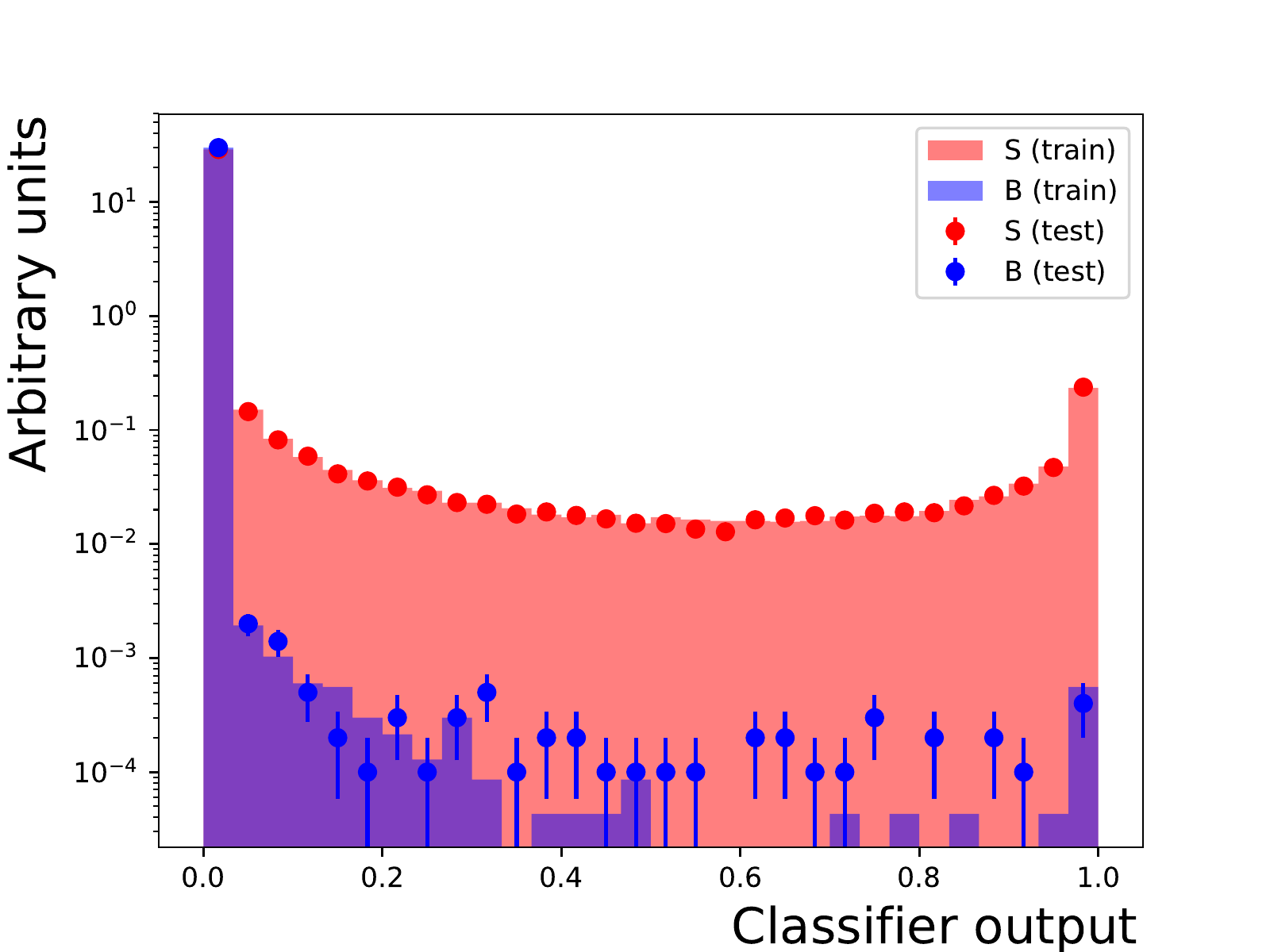}~
  }
  \caption{Histograms of the classifier outputs for the testing and training datasets of the
  value of the output neuron for single layer neural networks that are
  optimised with the three different loss functions. Both signal and
  background events have an equal weight.}
  \label{fig:discriminator}
\end{figure}

\begin{figure}[t]
\centering
\subfigure[Binary cross entropy]{
    \includegraphics[width=0.5\textwidth]{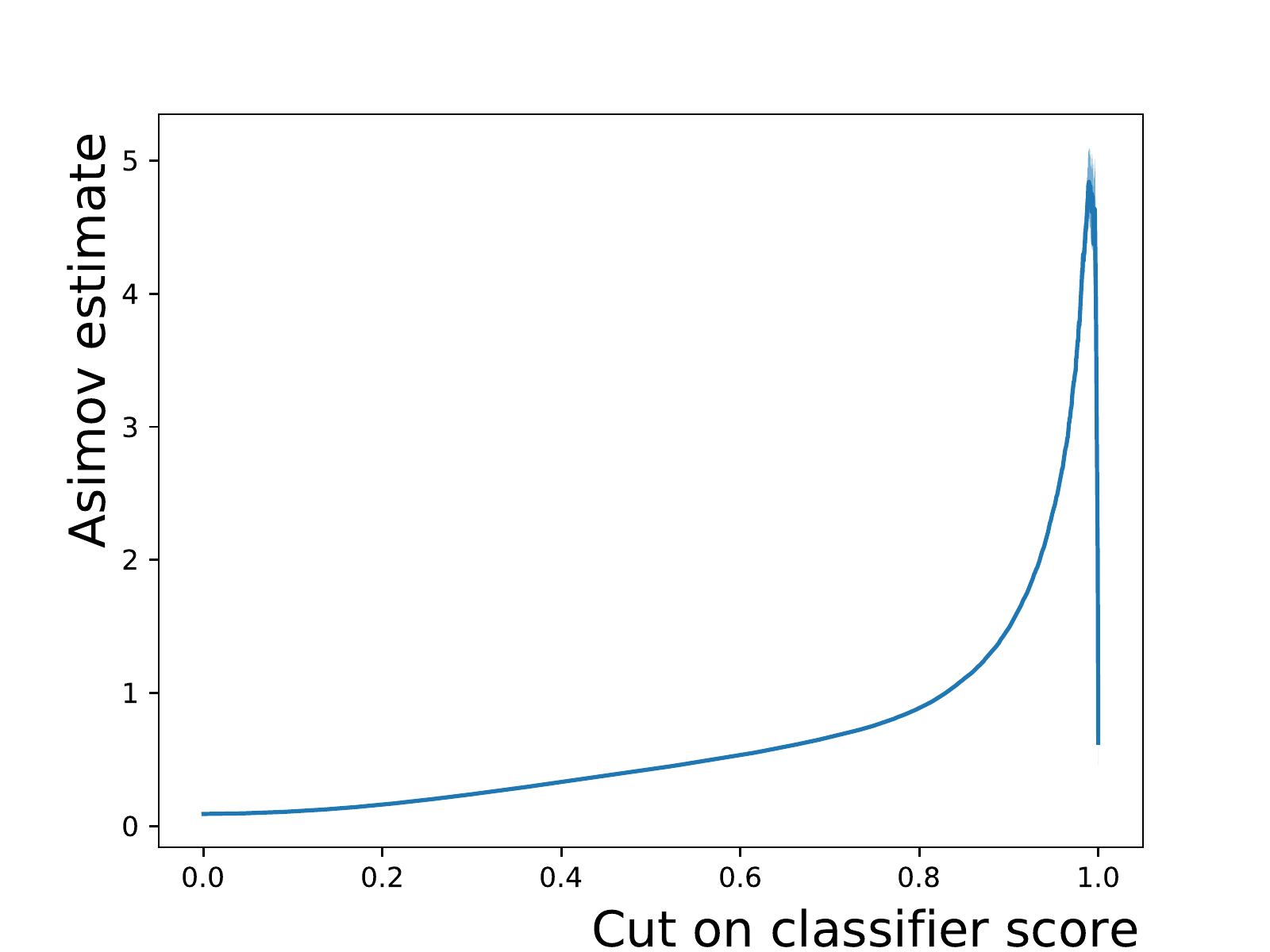}~
  }~
  \subfigure[$\ell_{s/\sqrt{s+b}}$]{
\includegraphics[width=0.5\textwidth]{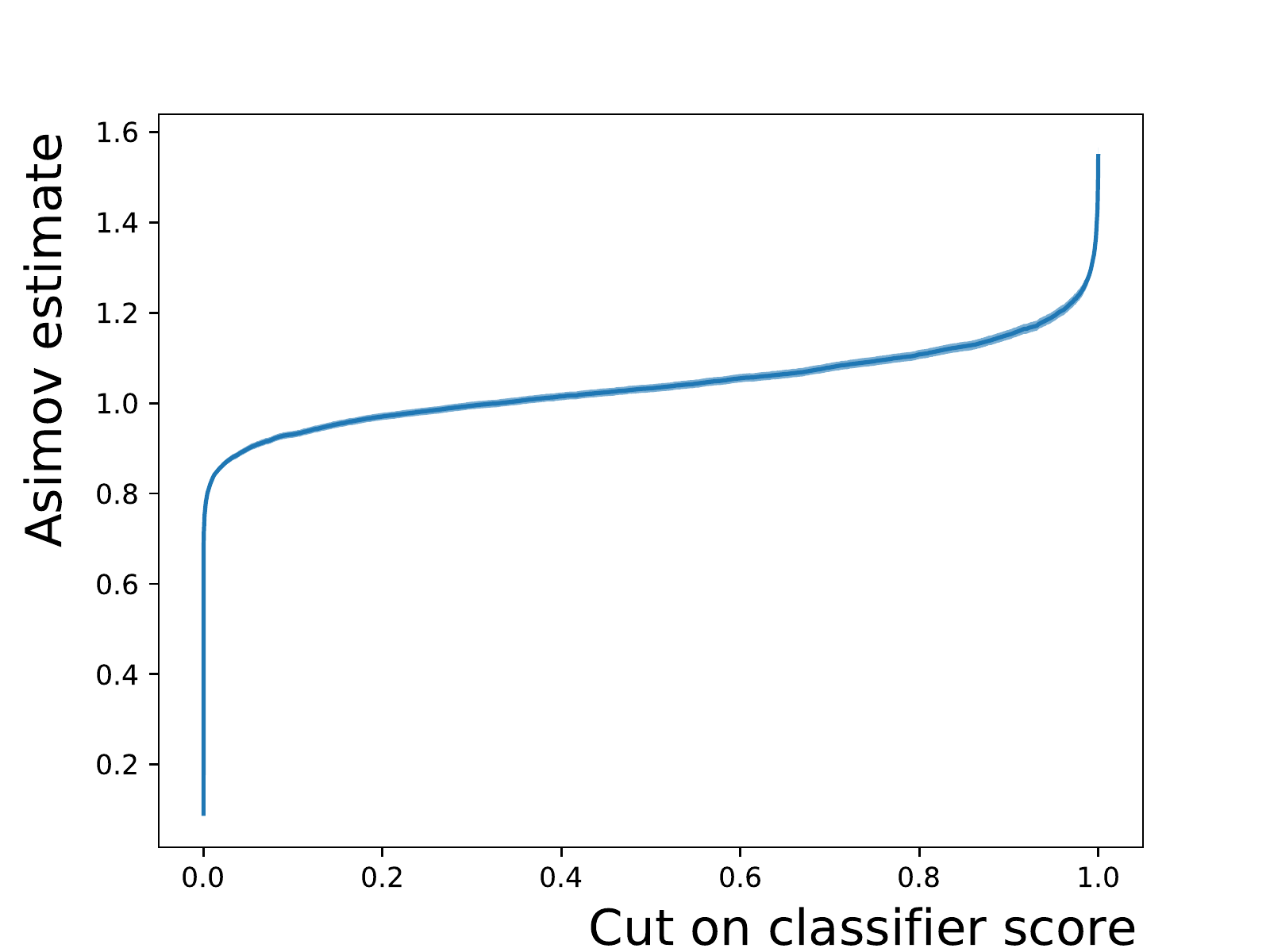}
  }\\
  \subfigure[$\ell_{Asimov}$]{
\includegraphics[width=0.5\textwidth]{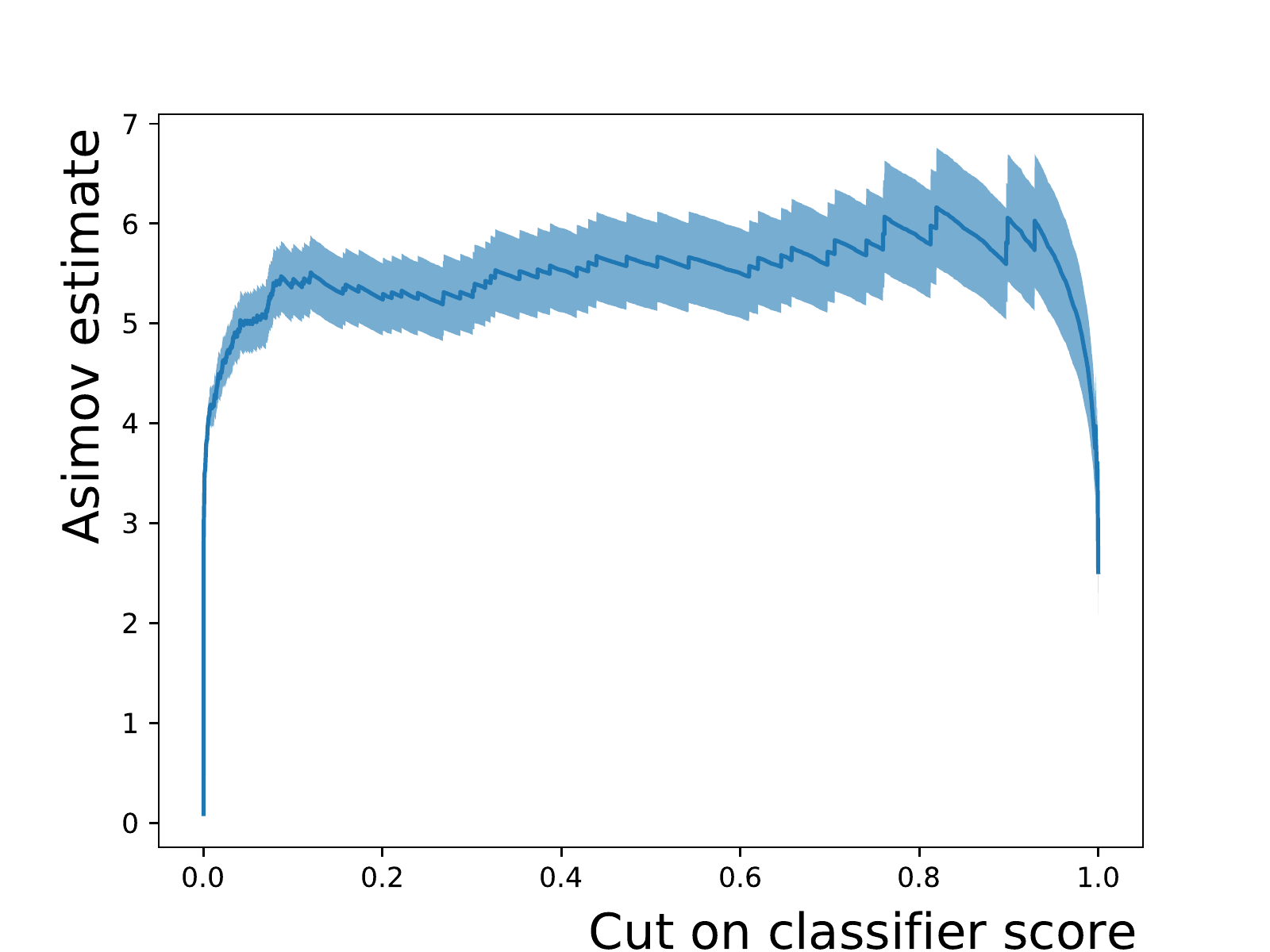}~
  }
  \caption{ The Asimov estimate of significance with a $50\%$ systematic
  uncertainty on the background as a function of classifier cut for
  the one layer network trained with the three different loss
  functions. This is calculated for the compressed SUSY mass point. A
  $\pm1\sigma$ error band of the estimate based on the simulation
  statistics is included in light blue. The significance is calculated
  with both signal and background events weighted to reflect the
  number of expected events observed in $30$~fb$^{-1}$ of $14$~TeV
  proton-proton collisions.}
  \label{fig:significanceCut}
\end{figure}

\begin{table}
  \centering
\begin{tabular}{ cccccccccc }
   \multirow{2}{*}{Loss} & \multicolumn{3}{c}{$Z_A(\sigma_b/b = 0.1)$} & \multicolumn{3}{c}{$Z_A(\sigma_b/b = 0.3)$} & \multicolumn{3}{c}{$Z_A(\sigma_b/b = 0.5)$} \\
                    & $s$ & $b$ & $\sigma$ & $s$ & $b$ & $\sigma$ & $s$ & $b$ & $\sigma$\\
\hline
  \multicolumn{10}{c}{Uncompressed model, $m_{stop}=900\GeV$,
  $m_{LSP}=100\GeV$ }\\
\hline
   Cross entropy& $8.7$ & $1.7$ & $4.5\pm0.3$& $7.7$ & $1.2$ & $4.0\pm0.3$&
  $7.7$ &  $1.2$ &$3.5\pm0.3$\\
   $\ell_{s/\sqrt{s+b}}$& $7.7$ & $1.3$ & $4.3\pm0.3$& $7.7$
  & $1.3$ & $3.9\pm0.3$ & $7.7$ & $1.3$ & $3.4\pm0.3$\\
   $\ell_{Asimov}$& $6.6$ & $1.6$ & $3.6\pm0.2$& $3.5$ &
  $0.1$ & $4.0\pm0.5$& $3.3$ & $0.1$ & $4.2\pm0.7$\\
\hline
  \multicolumn{10}{c}{Compressed model, $m_{stop}=600\GeV$,
  $m_{LSP}=400\GeV$ }\\
\hline
  Cross entropy& $74.4$ & $18.2$ & $10.7\pm0.3$& $44.0$ & $7.7$ & $6.8\pm0.3$&
  $40.5$ & $6.8$ & $4.8\pm0.3$\\
  $\ell_{s/\sqrt{s+b}}$& $323$ & $326$ &$7.0\pm0.1$ & $323$
  & $326$ & $2.56\pm0.03$& $324$ & $327$ & $1.55\pm0.02$\\
  $\ell_{Asimov}$& $78.4$ & $19.4$ & $10.8\pm0.3$& $25.9$ &
  $3.2$ & $6.8\pm0.4$& $11.9$ & $0.5$ & $6.2\pm0.6$\\
\\
\end{tabular}
\caption{The expected number of signal, $s$, and background, $b$,
  events that lead to the best significance, $\sigma$, gained with a classification of
  signal and background obtained with a cut on the output neuron of a
  single layer neural network optimised with the three different loss
  functions. $Z_A$ is calculated with a $10\%$,$30\%$ or $50\%$
  systematic uncertainty
  on the background. The expected numbers are calculated given 
  30~fb$^{-1}$ of 14~TeV LHC data.}
\label{tab:results}
\end{table}

Optimisation is carried out with three different loss functions,
$\ell_{s/\sqrt{s+b}}$, $\ell_{Asimov}$ (with a range of systematic
errors tested, from $5$-$50\%$ of the background counts) and the binary
cross entropy. The {\it Adam} optimisation algorithm is used to
minimise the loss \cite{ADAM}. As discussed in Sec.~\ref{sec:NN}, a
size of 4096 is used for the new loss functions. A batch size of 128
was found to work well for the binary cross entropy and is used
throughout the studies. Each of the networks are trained until the value of the
loss function in the test data does not decrease for two epochs.

The evolution of $\ell_{s/\sqrt{s+b}}$ and accuracy as a function of epoch
for the test and train datasets can be seen in
Fig.~\ref{fig:lossaccevolution}. It is evident that the minimisation
of the loss function does not directly correspond to a monotonic
increase in the accuracy, as would be the case for the binary cross
entropy. The reason for this can be clearly seen in
Fig.~\ref{fig:discriminator}, which shows the value of the output
neuron for the test and train datasets, split into signal and
background events. The new loss functions maintain the purity of the
signal classification at the expense of signal events, which are
misclassified as background. The binary cross entropy, however, puts
equal weight into the correct classification of the signal and
background. This allows the new loss functions to select regions of phase space
that have a purer signal to background ratio when it is advantageous
to the overall significance to do so. 

The performance for the two separate SUSY mass points introduced in
Sec.~\ref{sec:stop} are considered. A separate training is carried out
for each loss function for both of the models. To evaluate their
performance, the value of the output neuron for each network, the
classifier score, is considered. Potential cuts on this score are
scanned through, where all the events that have a score greater than a cut
value are kept. After these cuts the Asimov estimates of the
significance for a variety of systematic uncertainties are calculated.
Plots for each loss function, showing the significances as a function of cut value for the
compressed SUSY model with a systematic uncertainty of $50\%$ of the background
counts, is displayed in Fig.~\ref{fig:significanceCut}. The maximum
significances obtainable with this method, along with
the expected number of signal and background events remaining after
the chosen cut, for each of the loss function
and SUSY models are presented in Tab.~\ref{tab:results}. The error on
the Asimov estimate is calculated by propagating the Poisson
uncertainty of the background and signal counts.

For the uncompressed model, all three of the loss functions perform
similarly. In this case there are appreciable differences in the
distributions of the input variables between the signal and background
samples, but a low number of signal events expected.
This results in a significance that is limited by the statistical
uncertainty, rather than the systematic uncertainty. As there is no
advantage to factoring in the systematic uncertainties in this case, each loss function manages to find an
optimal solution. However, there is a difference in purity of the
selected signal and background events for the different loss
functions. The maximum value of $s/b$ for the binary cross entropy
optimisation, as
a function of cut on the classifier, is $\sim6$, whereas when
optimising with $\ell_{Asimov}$ an $s/b$ of $\sim30$ is obtained. The
optimisation of
$\ell_{s/\sqrt{s+b}}$ also resulted in an $s/b$ of
$\sim6$. This
suggests that when the optimisation is aware of the systematic error,
there is pressure for it to select a different region of
phase spaced with a higher purity of signal events. 

When optimising the networks for the compressed model, there is a
bigger difference in the performance. In this case the signal and
background distributions are more similar, but the cross section is
higher, so there is a larger number of signal events for the network to
correctly classify. This makes the result much more sensitive to
systematic uncertainties than statistical uncertainties. The only optimisation that is aware of the
systematic uncertainty is the one with $\ell_{Asimov}$. This results in a better
performance for this loss, especially when considering higher
systematic uncertainties. This gain is due to optimisation with
$\ell_{Asimov}$ resulting in a
higher purity. The optimisation with
$\ell_{s/\sqrt{s+b}}$ contains no information about the systematic
uncertainty and performs most poorly, finding a maximum
$s/b$ of only $\sim 1$.
The binary cross entropy still performs well, but requires a
cut on values of the classifier score very close to one, interpretable
as a very high probability of signal event and low probability of a
background event.  One other noticeable feature of the new loss
functions is that it is unnecessary to tune the cut on the classifier to gain the optimal
performance, all cuts above $\sim 0.5$ perform well.

A similar study was carried out considering 300~fb$^{-1}$, rather than
30~fb$^{-1}$,
of LHC data. In this case the improvement shown by $\ell_{Asimov}$ was
starker, with a $9.9\pm1.6~\sigma$ separation achieved with a
systematic uncertainty of $50\%$. By comparison, the binary cross
entropy achieved a 
$6.4\pm0.7~\sigma$ separation.

\section{Conclusions} 

In searches for new physics at particle colliders, analyses aim at
maximising the statistical significance of signal events over the
background in a given search region. Neural networks offer a powerful
technique for classifying signal and background events and thus
determining search regions. New loss functions that directly maximise
two estimations of statistical significance, $s/\sqrt{s+b}$ and the
Asimov estimate, have been introduced. The optimal batch size is
required to be fairly large, $O(10^3)$, to minimise 
statistical fluctuations 
on the estimation of the significance in the loss functions. When optimising
with $\ell_{Asimov}$ it is also found that some pretraining with
$\ell_{s/\sqrt{s+b}}$ results in a significant speed up of the
optimisation time. 

These strategies have been
tested in a simulation of a SUSY search for direct stop production in
$30$~fb$^{-1}$ of $14$~TeV proton-proton collisions at the LHC. The
optimisation with
$\ell_{Asimov}$ has been demonstrated to select signal events with a
higher purity than the standard loss function for neural network classification
tasks, the binary cross entropy. This can result in a better discovery
significance especially in models that are dominated by systematic
uncertainties.

This toy study provides a proof of principle of the loss functions
described in this paper, future studies will be carried out to
determine their usefulness in a more realistic analysis scenario. 

Examples of an implementation of the loss functions with the Keras
python library~\cite{chollet2015keras} can be found in the
\href{https://github.com/aelwood/hepML}{GitHub} repository containing
the code used for this study\footnote{\url{https://github.com/aelwood/hepML/blob/master/MlFunctions/DnnFunctions.py}}.

\acknowledgments
We would like to thank Marco Costa from the Scuola
Normale Superiore/University of Pisa and Leonid Didukh
for producing the simulation required for the studies
presented in this paper. We would also like to thank
Christian Contreras Campana, Isabell
Melzer-Pellmann and Olaf Behnke for valuable discussion in the process
of developing and writing up this work.
\clearpage

\bibliographystyle{utphys}
\bibliography{lit}

\end{document}